\documentclass[manuscript]{aastex6}

\usepackage{amsmath}
\usepackage{color}
\usepackage{comment}
\usepackage{url}
\usepackage{graphicx}

\begin{document}

\author{Kosuke Namekata\altaffilmark{1}}
\author{James R. A. Davenport\altaffilmark{2}}
\author{Brett M. Morris\altaffilmark{2, 3}}
\author{Suzanne L. Hawley\altaffilmark{2}}
\author{Hiroyuki Maehara\altaffilmark{4, 5} }
\author{Yuta Notsu\altaffilmark{6, 7, 12} }
\author{Shin Toriumi\altaffilmark{8} }
\author{Kai Ikuta\altaffilmark{1}}
\author{Shota Notsu\altaffilmark{9, 12} }
\author{Satoshi Honda\altaffilmark{10}}
\author{Daisaku Nogami\altaffilmark{1} }
\author{Kazunari Shibata\altaffilmark{11}}


\altaffiltext{1}{Department of Astronomy, Kyoto University, Kitashirakawa-Oiwake-cho, Sakyo, Kyoto 606-8502, Japan; namekata@kusastro.kyoto-u.ac.jp}

\altaffiltext{2}{Astronomy Department, University of Washington, Seattle, WA 98119, USA}

\altaffiltext{3}{Center for Space and Habitability, University of Bern, Gesellschaftsstrasse 6, 3012 Bern, Switzerland}

\altaffiltext{4}{Okayama Branch Office, Subaru Telescope, National Astronomical Observatory of Japan, NINS, Kamogata, Asakuchi, Okayama 719-0232, Japan}

\altaffiltext{5}{Okayama Observatory, Kyoto University, 3037-5 Honjo, Kamogata, Asakuchi, Okayama 719-0232, Japan}

\altaffiltext{6}{Laboratory for Atmospheric and Space Physics, University of Colorado Boulder, 3665 Discovery Drive, Boulder, CO 80303, USA }

\altaffiltext{7}{National Solar Observatory, 3665 Discovery Drive, Boulder, CO 80303, USA}

\altaffiltext{8}{Institute of Space and Astronautical Science (ISAS), Japan Aerospace Exploration Agency (JAXA), 3-1-1 Yoshinodai, Chuo-ku, Sagamihara, Kanagawa 252-5210, Japan}


\altaffiltext{9}{Leiden Observatory, Leiden University, P.O. Box 9513, 2300 RA Leiden, The Netherlands}

\altaffiltext{10}{Nishi-Harima Astronomical Observatory, Center for Astronomy, University of Hyogo, Sayo, Sayo, Hyogo 679-5313, Japan.}

\altaffiltext{11}{Astronomical Observatory, Kyoto University, Kitashirakawa-Oiwake-cho, Sakyo, Kyoto 606-8502, Japan.}

\altaffiltext{12}{JSPS Overseas Research Fellow}

\title{Temporal Evolution of Spatially-Resolved Individual Star Spots on a Planet-Hosting Solar-type Star: Kepler 17}
\shorttitle{Temporal Evolution of Individual Star Spots on Kepler 17}

\shortauthors{Namekata et al.}

\begin{abstract}
Star spot evolution is visible evidence of the emergence/decay of the magnetic field on stellar surface, and it is therefore important for the understanding of the underlying stellar dynamo and consequential stellar flares.
In this paper, we report the temporal evolution of individual star spot area on the hot-Jupiter-hosting active solar-type star Kepler 17  whose transits occur every 1.5 days. 
The spot longitude and area evolution are estimated (1) from the stellar rotational modulations of \textit{Kepler} data and (2) from the brightness enhancements during the exoplanet transits caused by existence of large star spots.
 As a result of the comparison, number of spots, spot locations, and the temporal evolution derived from the rotational modulations is largely different from those of in-transit spots.
We confirm that although only two light curve minima appear per rotation, there are clearly many spots present on the star.
We find that the observed differential intensity changes are  sometimes  consistent with the spot pattern detected by transits, but they sometimes do not match with each other. 
Although the temporal evolution derived from the rotational modulation differs from those of in-transit spots to a certain degree, the emergence/decay rates of in-transit spots are within an order of magnitude of those derived for sunspots as well as our previous research based  only on rotational modulations. 
This \textcolor{black}{supports} a hypothesis that the emergence/decay of sunspots and extremely-large star spots on solar-type stars occur through the same underlying processes.

\end{abstract}
\keywords{starspots -- stars: solar-type -- stars: activity -- sunspots}

\section{Introduction} \label{sec:int}
The Sun and other solar-type stars show similar magnetic activity.
In the case of some solar-type stars, very high magnetic activities such as large star spots, superflares, and high X-ray activities have been reported by many authors \citep[e.g.,][]{2005LRSP....2....8B,1997ApJ...483..947G,2012Natur.485..478M,2017ApJ...851...91N}.
They are expected to share the same underlying process, though the evidence is quite limited.
 There is a strong and increasing interest in the solar-stellar connection that arises from the question of effect of such extreme phenomena on exoplanet habitability around active stars \citep[e.g.,][]{2010AsBio..10..751S,2016NatGe...9..452A,2017ApJ...848...41L} as well as from the curiosity about possible extreme events on the Sun \citep[e.g.,][]{2013A&A...549A..66A,2013PASJ...65...49S,2017ApJ...850L..31H}. 

Star spots have been investigated as the visible manifestations of the magnetic fields on the stellar surfaces, and the stellar magnetic properties such as occurrence frequencies 
\citep[e.g.,][]{2017PASJ...69...41M,2019ApJ...876..58N}, 
stellar cycles \citep[e.g.,][]{2005AN....326..283B}, and differential rotations \citep[e.g.,][]{2016MNRAS.461..497B,2013A&A...560A...4R} have been estimated based on the star spot signatures.
Investigations on the temporal evolution of individual star spots are at present a major challenge but quite important for several kinds of fields.
(i) Spots are essential for the occurrence of flares, not only on the Sun \citep[see,][and references therein]{2019arXiv190412027T} but also on the stars \citep[e.g.,][]{2013PASJ...65...49S}, thus the flux emergence on stellar surface is a key to understand why and how stellar superflares occur.
(ii) The decay pattern of spots is thought to be related to the convective diffusion, which can constrain the stellar diffusion coefficient parameters on performing numerical modeling of the stellar dynamo.
(iii) It is also important for the determination of stellar rotational periods because most of the studies implicitly assume that the rotational period is much smaller than the spot lifetimes \citep{2014ApJS..211...24M,2018ApJ...868..151D}.

Up to now, the temporal evolution of the individual star spots has also been investigated using the long-term ground-based and space-based photometry.
In the 1990s, the temporal evolution of star spots area on the RS CVn-type stars, M, K-type stars, and young stars was investigated by tracing the local minima of the stellar light curve \citep{1994IAPPP..55...51H,1994A&A...282..535S}.
The developments of Doppler Imaging techniques indicated long-lived polar spots on active young stars \citep{1999A&A...347..212S,2012A&A...548A..95C} and made it possible to measure the variation rates of star spots on a sub-giant star \citep{2015A&A...578A.101K}.
Also, the developments of light curve modeling and inversion codes have been carried out \citep[see,][]{2008AN....329..364S,2009A&ARv..17..251S,2012A&A...543A.146F}; such the applications show varying evolution of star spots on solar-type stars \citep{2012A&A...543A.146F,2014ApJ...795...79B}.
Later, \cite{2017MNRAS.472.1618G} developed a simple method to estimate the decay timescales of star spots on active stars, and showed that there is a positive correlation between the mean decay timescales and star spot area of the star, which is similar to the solar empirical relation \citep[so-called Gnevyshev-Waldmeier relation;][]{Gnevyshev,1955epds.book.....W}.

More recently, \cite{2019ApJ...871..187N} extended the simple local minima tracing method introduced by \cite{1994IAPPP..55...51H} to a huge amount of solar-type stars observed by \textit{Kepler} Space Telescope, and found 56 favorable cases of the individual star spot evolutions on the solar-type stars.
The results show some consistency between sunspots and star spots in the relations of the emergence/decay rates and spot area.
 \cite{2019ApJ...871..187N} selected only favorable spots separated from other spots in longitude, but still have the following problems:
(1) There would be large uncertainties due to several causes: stellar inclination, the extent of polar spotting, the extent of contamination from other spots at different positions, and the number of spots that local minima have \citep[see,][]{2018ApJ...865..142B,2018ApJ...863..190B,2019ApJ...871..187N}.
Also, the unspotted stellar brightness level is unknown in the \textit{Kepler} light curves \citep[e.g.,][]{2018ApJ...865..142B}.
To evaluate those effects, the development of and comparison with light curve modeling method are necessary, but even these suffer from severe degeneracies that make it unlikely to discover the true spot distribution only from a light curve.
(2) Moreover, since \cite{2019ApJ...871..187N} selected only favorable spots separated from other spots in longitude, so there can be selection biases.
Most light curves are highly disturbed by stellar differential rotations and some have long-lived active longitudes \citep[or active nests][]{1987ARA&A..25...83Z,2003A&A...405.1121B}, so the validation of our results is difficult on the other stars.
In addition, the criteria of \cite{2019ApJ...871..187N} can miss long-lived spots ($>$ 1 year) because of the stellar differential rotations and limited $Kepler$'s observational time ($\sim$ 4 years). 

Here we take an approach using exoplanet transits. 
During the transit, the additional small brightness variations can be sometimes detected which are caused by an exoplanet passing in front of dark spots on the stellar surface \citep{2003ApJ...585L.147S}.
By analyzing the brightness variations, the spot locations and area can be estimated with the spatial scale as small as the exoplanet diameter \citep{2015PhDT.......177D,2016ApJ...831...57E,2017ApJ...835..294V,2017ApJ...846...99M}.
Kepler 17 is a hot-Jupiter-hosting active solar-like star, and it is a good target to estimate the spot evolutions by using the transit data \citep{2015PhDT.......177D}.
In Kepler-17, the transit path of the hot Jupiter (Kepler 17b) is almost parallel to the rotational direction and the same spots can be detectable for many times.
Therefore, by using the transit of Kepler-17 data, we can estimate the temporal evolution of spatially resolved individual star spots. 
Because of this unique feature, Kepler 17, as well as its planet, has been well-studied \citep{2011ApJS..197...14D,2012A&A...PAR,2017ApJ...835..294V}.
\cite{2012A&A...Activity} performed the light curve inversion from the rotational modulation to the stellar surface filling factor, \cite{2016ApJ...831...57E} estimated the stellar magnetic cycles by using the transit data, and \cite{2017ApJ...835..294V} estimated the stellar differential rotation in this star.
Later, \cite{2019arXiv190404489L} compared the results of the transit method with those of the light curve inversion method, and discussed their consistency, especially the differential rotation and activity cycles.
\cite{2015PhDT.......177D} showed some examples of the temporal evolution of the spots in transit, but did not show the detailed comparisons with sunspots.\footnote[1]{In \cite{2019ApJ...871..187N}, we have mistakenly cited \cite{2014ApJ...795...79B} as an example of the estimations of star-spot lifetime by the transit method, but the work of \cite{2014ApJ...795...79B} was based on \cite{2012A&A...Activity}, and they analyze only the rotational modulations of Kepler 17.  Here, we revise the description.}

In this paper, we discuss the temporal evolution of spatially resolved star spots in transit on Kepler 17 based on the work by \cite{2015PhDT.......177D} in order to confirm the relations between sunspots and star spots discussed by \cite{2019ApJ...871..187N}.
Because long-lived active longitudes
have been reported by \cite{2019arXiv190404489L}, the star is also a good target to confirm to what extent we can measure the spot area and variation rates of individual spots based on the $Kepler$'s rotational light curve.
We analyzed the rotational modulations of Kepler 17 by using the local minima tracing method \citep{2019ApJ...871..187N} and newly-developed light curve modeling method \citep{Ikuta2019}, and compared the results with the spots detected in-transit.
The light curve modeling method is developed for the validation of the local minima tracing method.
In Sect. \ref{sec:data}, we introduce the data set and the stellar parameters of Kepler 17 and Kepler 17b.
In Sect. \ref{sec:ana}, we show our analysis methods: (a) local minima tracing method, (b) light curve modeling method, and (c) transit method.
In Sect. \ref{sec:re}, we show the results of our analysis, and in Sect. \ref{sec:dis} we discuss the results.

\section{Stellar Parameters and Data} \label{sec:data}

Kepler 17 is a solar-like main sequence star with spectral type G2V.
Its mass is 1.16 $\pm$ 0.06 $M_\odot$, radius is 1.05 $\pm$ 0.03 $R_\odot$, effective temperature is 5780 $\pm$ 80 K \citep{2011ApJS..197...14D,2012A&A...PAR,2017ApJ...835..294V}, and the stellar age is estimated to be less than 1.78 Gyr \citep{2012A&A...PAR}.
The star has rotational brightness variations with its star spots whose rotational period is about 12.4 days \citep{2012A&A...PAR}.
The large amplitude of these brightness variations show that this star has large star spots covering about 7\%
of the surface \citep{2019arXiv190404489L}, which is much larger than the observed maximum sizes of sunspots.

In the Kepler-17 system, a hot-Jupiter was first detected with \textit{Kepler} space telescope.
The planet mass is 2.45 $\pm$ 0.01 $M_{Jup}$, planet radius is 0.138 $\pm$ 0.001 $R_{\rm star}$, and orbital period is 1.4857108 days \citep[][]{2011ApJS..197...14D,2012A&A...PAR,2017ApJ...835..294V}.
During the exoplanet transits, small brightness enhancement can be observed, and this is thought to be caused by the exoplanet hiding the star spots on the stellar photosphere.
By modeling the transit light curve, individual star spot sizes and locations can be estimated with much smaller spatial resolution \citep[$\sim R_{\rm plant}/R_{\rm star} \sim 0.1 \sim 20^{\circ}$, see e.g.][]{2015PhDT.......177D,2017ApJ...846...99M,2017ApJ...835..294V} than  the rotational modulation methods \citep[$\sim 100^{\circ}$,][]{2018ApJ...863..190B}. 
Notable features of this system are as follows: (1) the inclination angle of the star is $\sim$ 90$^\circ$; (2) exoplanet path is almost perpendicular to the stellar rotational axis ($\sim$ 89$^\circ$), and the projected latitude beneath the exoplanet transit chord is near the equator ($\sim$ -4.6$^\circ$); and (3) the transits occur every 1.5 days (significantly shorter than stellar rotational period $\sim$ 12 days).
Thanks to these unique features, the same star spots can be detectable several times within the one rotational period, and even the recurrent spots can be traced over time.
Therefore, Kepler 17 is the best target to measure the temporal evolutions of the individual star spot areas \citep{2015PhDT.......177D}.

For the rotational modulations, we used all the available \textit{Kepler} pre-search data conditioning (PDC) long-cadence ($\sim$30 min) data \citep[\textit{Kepler} Data Release 25,][]{2016KeplerDR25}, in which instrumental effects are removed.
As for the transit, we used PDC short-cadence data.
Since the long-term light curve modeling is very expensive, we only analyzed the long-cadence data from Quarter 4 to 6, when the short cadence mode began to observe Kepler 17.
The in-transit star spot analysis method was applied to only observations taken at short cadence.

\section{Analysis Method} \label{sec:ana}
We investigate temporal evolutions of star spots on Kepler 17 by using the rotational modulations with simple traditional local minima tracing modeling \citep[hereafter local minima tracing method; see Sect. \ref{sec:ana1};][]{1994IAPPP..55...51H,2019ApJ...871..187N} and an MCMC-based simple light curve modeling \citep[hereafter light curve modeling method; see Sect. \ref{sec:ana2};][]{Ikuta2019}.
We also measured the temporal evolution of the starspots by transit modeling with the {\sf STSP} code \citep[hereafter in-transit method; see Sect. \ref{sec:ana3};][]{2015PhDT.......177D,2017ApJ...846...99M}.

\subsection{Method I: Local Minima Tracing Method} \label{sec:ana1}
The first method is a local minima tracing method which is firstly proposed by \cite{1994IAPPP..55...51H} to measure the temporal evolutions of star spots.
In this method, each star spot can be identified by the repetition of the local minima over the rotational phases.
The light curve of a rotating star with star spots shows several local minima when the spots are on the visible side \citep{2013ApJ...771..127N,2017PASJ...69...41M}. 
By tracing the local minima with the rotational period of the star, we can estimate the lifetimes of the large-scale spot patterns. 
We can estimate the temporal evolution of star spot coverage by measuring the local depth of the local minima from the nearby local maxima as an indicator of spot area \citep[hereafter local-minima spot; see][for the detailed method]{2019ApJ...871..187N}.
We first obtained the smoothed light curve by using locally weighted polynomial regression fitting \citep[LOWESSFIT;][]{LOWESS} to remove flare signature and noise.
In the LOWESSFIT algorithm, a low degree polynomial is fitted to the data subset by using weighted least squares, where more weight is given to nearby points. 
We used the $lowess$ function incorporated in the \texttt{statsmodels} $python$ package.
We detected the local minima as downward convex points of the smoothed light curve; i.e., the smoothed stellar fluxes $F(t)$ satisfy $F (t_{(\rm n-m)})$ $\rm <$ $F (t_{(\rm n-m-1)})$ and $F (t_{(\rm n+m)})$ $\rm <$ $F (t_{(\rm n+m+1)})$.
Here, $m$ takes a value of [0, 1, 2], $t$ is time, and $n$ is time step.
When we estimate the spot area from the local depth of the light curve, we assume that the spot contrast is 0.3.
The advantage of this method is its low computational cost, so it is suitable for statistical analyses, as discussed in \citet{2019ApJ...871..187N}.
This star is a good candidate to evaluate how the simple method can estimate the temporal evolution of star spot areas by comparing with the other methods.

\subsection{Method II: Light Curve Modeling Method} \label{sec:ana2}

Light curve modeling methods for a rotating, spotted star have been developed by several authors \citep[e.g.,][]{2006PASP..118.1351C,2012A&A...543A.146F,2014A&A...564A..50L}.
Here, we also developed a light curve modeling method under some simple assumptions \citep[see][in prep. for details]{Ikuta2019}.
The spot contrast is assumed to be constant (=0.3) because it is highly degenerate with spot area \citep[][]{2013ApJS..205...17W}.
We are interested in the spot evolution, so each spot area is assumed to emerge and decay linearly for simplicity. 
The real spot evolution may be more complex, but it will not give a significant effect on our results because we will focus on the relations of spot maximum size and mean variation rate as presented by \cite{2019ApJ...871..187N}.
A more realistic spot evolution model should be done in our future works.

\begin{figure*}[htbp]
\begin{center}
\includegraphics[scale=0.5]{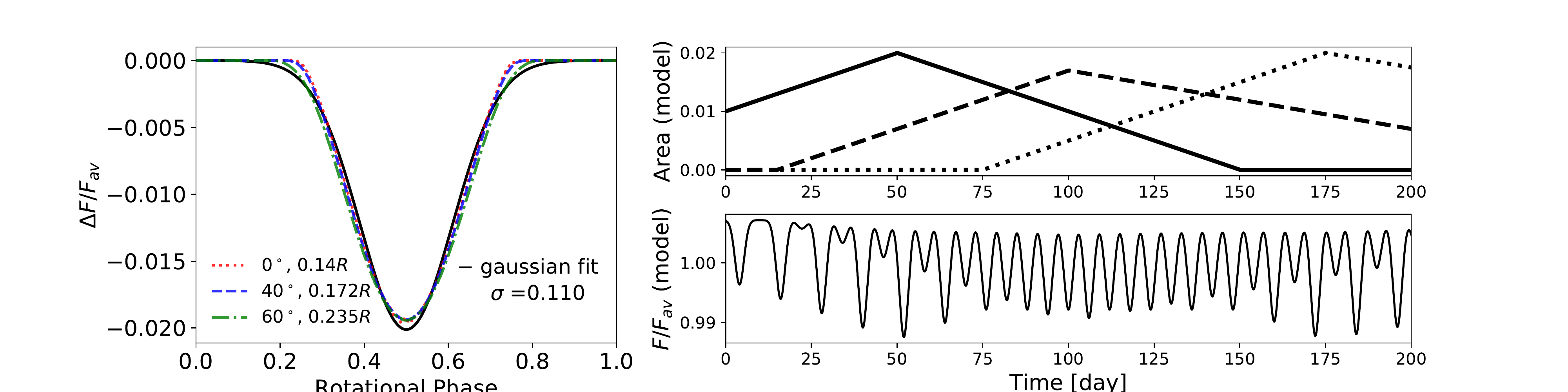}
\caption{This figure shows a detailed explanation of our model of out-of-transit rotational modulation. 
Left: light curves of a rotating star with (red) a spot of 0.14 $R_{\rm star}$ on the equator of stellar surface, (blue) a spot of 0.172 $R_{\rm star}$ on 40$^\circ$ and (green) a spot of 0.235 $R_{\rm star}$ on 60$^\circ$. 
The black solid line behind is the model light curve that we use in our MCMC modeling, which is derived by fitting the mean of the color lines with a gaussian function. Upper right: one model of the temporal evolution of star spot. The values of the vertical axis is just a fraction in the light curves ($A(t)$ in the equation of Equation \ref{eq:model}). Lower right: the light curve generated from the upper temporal evolutions by using our gaussian model.}
\label{fig:model_ReMC}
\end{center}
\end{figure*}

In the left panel of Figure \ref{fig:model_ReMC}, modeled stellar light curves are plotted in several cases of spot size and latitude.
In the model, we use the stellar surface model separated with grids by assuming linear-limb darkening \citep[see][]{2013ApJ...771..127N}.
As one can see, the stellar light curve of a low-latitude small spot  is highly degenerate with  those of high-latitude large spots within the \textit{Kepler} photometric errors of  0.1\% \citep[see also][]{2013ApJS..205...17W}.  
By considering this, we did not consider the latitude information, so there should be uncertainty in area estimations caused by the projection effect at higher latitudes.

There are analytical models which reproduce stellar light curves from the spot parameters \citep[e.g.,][]{2012MNRAS.427.2487K}.
In this study, we adopt an approach to use a Gaussian function to approximate stellar rotational light curves.
As in Figure \ref{fig:model_ReMC}, all of the light curve in the different cases show similar light curves in Kepler 17, which can be fitted by a Gaussian function as indicated in black line, with the Kepler's photometric errors ($\sim$ 0.1 \%).
We used the following function as a standard light curve:
\begin{eqnarray}
\Delta F(t) = -A \times \rm exp\left( -\frac{(\it t-t_{\rm 0})^{\rm 2}}{(\sigma P_{\rm rot})^{\rm 2}}\right), 
\end{eqnarray}
where $A$ is spot area, $\sigma$ is non-dimensional factor (=0.110) derived by the Gaussian fitting, $t$ is time, $t_0$ is the standard time when the spot are on the visible side, and $P_{\rm rot}$ is the rotational period of a given spots.
The right of Figure \ref{fig:model_ReMC} is an example of the application of this Gaussian light curve to the multi-spot case.
In this case, the model light curve $\Delta F(t)$ can be obtained as
 \begin{eqnarray}  \label{eq:model}
\Delta F(t) = \sum_{n=1}^{\rm N_{spot}} \sum^{\infty}_{m = -\infty} \left(-A^{\rm n}(t) \times \rm exp\left(-\frac{(\it t-t^{\rm n}_{\rm 0}+ m \it P^{\rm n}_{\rm rot})^{\rm 2}}{(\sigma P^{\rm n}_{\rm rot})^{\rm 2}}\right) \right).
\end{eqnarray}
Here, $A(t)$ is a spot area as a function of time $t$, and $n$ is the spot number.
Note that our original code described by \cite{Ikuta2019} can generate a more realistic light curve considering the inclination and spot latitude.

Here, the following six parameters are necessary for each spot: maximum area, emergence rate, decay rate, maximum timing, standard longitude ($t_0$), and rotational period ($P_{\rm rot}$).
The total number of parameters is six times the number of spots (here, the number of spots is set to be five). 
We carried out a parameter search to estimate the most-likely parameters well reproducing the observed light curve 
with Markov chain Monte Carlo (MCMC). 
MCMC methods have become an important algorithm in not only astronomical \citep{2017ARA&A..55..213S} but also various scientific fields \citep[e.g.,][]{ApplicationMCMC2001}.
It can generate samples that follow a posterior distribution by selecting the sampling way based on the likelihood function $L$ in the Bayesian theorem.
Here, we take the likelihood function as a product of Gaussian function on each time:
\begin{eqnarray}
L = \prod_{i=1}^{N_{\rm data}} \it \frac{\rm 1}{\sqrt{{\rm2}\pi \sigma_{\rm 0}^{\rm 2}}} \rm exp\left(\it - \frac{\left(\left(\frac{F(t_i)}{F_{\rm av}}\right)_{\rm obs} - \left(\frac{F(t_i)}{F_{\rm av}}\right)_{\rm model} \right)^{\rm 2}}{{\rm 2}\sigma_{\rm 0}^{\rm 2}}  \right).
\end{eqnarray}
We adopt the traditional Metropolis-Hastings algorithm \citep{1953JChPh..21.1087M,MH1970}.
We also adopt the Gaussian function as the proposal distribution, and the step length (proposal variance) of the Gaussian was adaptively-tuned for each step by considering the acceptance ratio of MCMC sampling converged to be 0.2 \citep[so-called adaptive MCMC algorithm, e.g.,]{Gilks1998,ArakiPT2013,2017ifs..confE..30Y}.
A uniform distribution is adopted as a prior distribution.

In our case, when sampling from a multi-modal distribution with a large number of parameters, the chains theoretically can converge to the posterior distribution, but practically seapking, the chains may not converge because of the limited sampling times. 
They can be trapped in a local mode for a very long time.
In order to avoid this, we apply the Parallel Tempering (PT) algorithm to our MCMC estimations \citep[e.g.][]{1996JPSJ...65.1604H,ArakiPT2013}.
The PT algorithm introduces auxiliary distribution by using the so-called inverse temperature ($\beta$), runs multiple MCMC samplings (hereafter ``replica'') simultaneously for each inverse temperature, and sometimes, exchange replica with a certain percentage.
The tempering procedure and exchange processes help samples to escape from a local maximum.
In practice, each Markov chain is controlled by the inverse temperature $\beta_i$ ($1=\beta_0 > \beta_1>\beta_2...>\beta_n>0$), and the auxiliary likelihood for each replica is expressed as $L_i^{\beta_i}$.
The higher-order replica has, therefore, the smaller valley to another local maximum than the lower-order one, and the parameters are easy to be searched across the wide range of the parameter space without being trapped in local maxima.
By repeating the exchange between the orders, the highest-inverse-temperature replica (= the target samples) can sample around the maximum likelihood within finite computation times.

We run the MCMC sampler for 500,000 steps after the exchanges no longer occur, and set the burn-in period as the first 100,000 steps.
As noted above, the replica change can occur theoretically even when sampling around the highest likelihood, but actually did not.
We finally take the parameter showing the highest likelihood value in the 400,000 steps, and the error bars are estimated from the posterior distributions.
For the parameter range, we take [0.001, 0.05] for maximum spot fraction ($A_{\rm max}$), [-5, -1] for emergence rate ($\Delta A$/day)  in log scale, [-5, -1] for decay rate ($\Delta A$/day) in log scale, [start time, end time] for maximum timing, [0$^\circ$, 360$^\circ$] for  standard Carrington longitude, and [11.63, 12.86] for rotational period ($P_{\rm rot}$).
 In this study, we adopt a four-spot model by considering the model evidence and its output. We also carried out five spot model and no significant new information was obtained.  
Initial input values are randomly selected in the parameter range and independently selected for each replica. 
The convergence was checked by visually checking the parameter changing for each step and marginalized posterior distribution with one parameter or two parameters.
 The MCMC method can estimate the parameters to best reproduce the light curve, but the reproduced spots   do not necessarily have much   to do with the spots that are actually present as we describe in the following paragraph.

\subsection{Method III: Transit Method ( {\sf STSP} code )} \label{sec:ana3}

We modeled the spots occulted during the exoplanet transit (hereafter in-transit spots) by using {\sf STSP} code\footnote[2]{The source code of {\sf STSP} can be seen here: \url{https://github.com/lesliehebb/stsp}} \citep{2015PhDT.......177D,2017ApJ...846...99M}.
The spot modeling approach applied in the present study is the same as that already described in Chap. 5 of \cite{2015PhDT.......177D} and L. Hebb et al. {\it in preparation}, to which we refer the reader for details. 
In brief, {\sf STSP} is a C code for quickly modeling the variations in the flux of a star due to circular spots, both in and out of the path of a transiting exoplanet. Static star spots are currently assumed in the code, and evolution of spot activity is typically analyzed by modeling many windows (e.g. individual transits) independently. 
{\sf STSP} also allows the user to fit light curves with arbitrary sampling for the maximum-likelihood spot positions and size parameters using MCMC, and has been used for modeling systems with no transiting exoplanets \citep{2015ApJ...806..212D}, as well as transiting systems with a range of geometries \citep[e.g.][]{2017ApJ...846...99M}. 
 Note that this application of MCMC algorithm is different from that described in the section \ref{sec:ana2} \citep[see,][]{2017ApJ...846...99M}.  
The MCMC technique allows us to properly explore the degeneracies between star spot positions and sizes. 
The modeled data is mostly taken from the work by \cite{2015PhDT.......177D}.
Please note that the model by \cite{2015PhDT.......177D} was originally carried out by assuming by the stellar radius of 1.1 $R_{\odot}$ and planet radius of 0.10 $R_{\rm star}$, which is slightly different from reported from the other study \citep[][]{2011ApJS..197...14D,2012A&A...PAR,2017ApJ...835..294V}.
This can lead to 10 \% errors in the estimated area and variation rates of area, but these do not significantly affect our results, because we will discuss much trends that are much larger than this difference.
For the {\sf STSP} outputs, we remove solutions of spots which are located on the stellar limb (i.e. distance from the disk center $>$ 1 - 2 $R_{\rm planet}$) and spots with the too large size (i.e., radius $>$ 0.25 $R_{\rm star}$) because the solutions in the ranges are likely to be biased.
 We also calculated the error bars of the estimated spot parameters based on the posterior distributions of MCMC samplings.

\section{Results} \label{sec:re}
\subsection{Local Minima Tracing Method in Comparison with Transit Reconstruction}\label{res:1}
Figure \ref{fig:lc:transit} shows the result of local-minima spots and a comparison with in-transit spots.
The middle panel shows the temporal evolution of local-minima spots.
As described by \cite{2019arXiv190404489L}, there are two active longitudes in this star as indicated with the different colors, and  spot group
Group A (red) is dominant compared to the spot group B (blue) in this period. 
 We use this notation because local light curve minima on the Sun are sometimes due to spot groups or ``active longitudes'' \citep[e.g.,][]{2005LRSP....2....8B,2019arXiv190404489L,2019ApJ...871..187N}. In general these terms may be misleading because the source of light curve minima might not be either of these phenomena. Minima can easily arise (even on the Sun) from a scattered configuration where more spots are on one hemisphere than the other. We have chosen to use conventional terminology, but one can more accurately read ``spot group A" as ``side A". 
The temporal evolution of the area of spot group A shows a clear emergence and decay phase, while spot group B does not.

The bottom panel of Figure \ref{fig:lc:transit} is a comparison of spot locations between local-minima spots and in-transit spots.
Two local minima are dominant in one rotational phase, while there are 5-6 spots are visible in the transit path.
The longitude of the local minima well matches those where in-transit spots are crowded, especially in the first 200 days (BJD 350 to 550+245833). 
 On the other hand, after  BJD 550+245833,  the red and blue circles lose track of most of the transit spots, and the longitude of local minima becomes delayed compared to the in-transit spots (i.e., the slower rotation period than that of the equator).
It  could   be due to the complex spot distribution changing, but  could  be caused by the new emergence of star spot on the higher latitude where spots are rotating slower because of the solar-like differential rotation of the star \cite[see, ][]{2019arXiv190404489L}.

Figure \ref{fig:validation_kepler17} indicates the comparison of the area of local-minima spots and in-transit spots.
The upper left panel shows a comparison between \textit{Kepler} light curve and the reconstructed one from the in-transit spots.
 Note that the reconstructed light curve in Figure \ref{fig:validation_kepler17} is subtracted by its mean value, and the zero level for the calculation of the reconstructed light curves is around ($F$-$F_{\rm av}$)/$F_{\rm av}$ $\sim$ 0.28. 
The differential amplitude of the observed light curve is comparable to or a little smaller than the one reconstructed from the spots in-transit.

The lower left panel of Figure \ref{fig:validation_kepler17} shows the comparison of temporal evolutions of the local-minima spot \textcolor{black}{area} (black), maximum \textcolor{black}{visible area of} in-transit spot (green), and total spot coverage in transit \textcolor{black}{path on half the star} (orange).
 The total spot coverage in the transit chord was calculated for the closest transits to the local minima. 
The right panel of Figure \ref{fig:validation_kepler17} is the comparison for each data point of the lower left panel of Figure \ref{fig:validation_kepler17}.
The area of local-minima spots is consistent with the maximum spot area in-transit.
The area of local-minima spots also have a positive correlation with the total spot area in transit, although the former is smaller by a factor of two.
This tendency can be also seen in the case of the Sun \citep{2018ApJ...865..142B}. 

\subsection{ Light Curve Modeling Method in comparison with Local-Minima-Tracing/Transit Method }\label{res:2}

Figure \ref{fig:ReMCSTSP} shows the results of light-curve-modeling spots in comparison with in-transit spots.
The upper panel shows a comparison between \textit{Kepler} light curve and the modeled one.
The \textit{Kepler} light curve is reproduced by our simple Gaussian model.
 That is because there is only hemispheric information in the light curve, which is straightforward to model (but not so accurate). 
The middle panel shows the estimated areal evolution of the light-curve-modeling spots.
The estimated parameters are listed on Table \ref{tab:mcmc}.
The bottom panel shows that the estimated location of the light-curve-modeling spots matches that of the in-transit spots, especially for the red and purple spots.
This would be because the purple and red spots are located near the equator.
On the other hand, green and blue spots are not rotating with the same rotational period as the equator, so there is no corresponding spot occulted by transit.
If we compare Figure \ref{fig:lc:transit} and \ref{fig:ReMCSTSP}, we can see that the locations of the local minima and the estimated spot are quite similar to each other.

 We also tried a five-spot model and no significant new information was obtained. 
Here, we note that long-term spot modeling, covering over a quarter ($\sim$ 90 days), should be done carefully because \textit{Kepler} light curves have an inevitable long-term instrumental trend in the light curve, and the absolute values may be unreliable.

\subsection{Comparison of Temporal Evolutions of Star Spot Area among Different Methods}\label{res:3}

We estimated the location and area of star spots occulted by the planet for all \textit{Kepler} short time cadence data (16 quarters in total).
The estimated result is plotted in Figure \ref{fig:dbscan}.
There are many long-lived recurrent spots that are located on the same longitude for a long time.
To pick up candidates of long-lived spots, we apply DBSCAN, a commonly-used data clustering algorithm, in the Python package \texttt{scipy} \citep{DBSCAN1996,2015PhDT.......177D}.
In brief, this algorithm finds the core points which have more than $N$ satellite points within the length of $\epsilon$, and find clusters by connecting the core points with each other.
The detected clusters are plotted with the colored symbols in Figure \ref{fig:dbscan}.
We interpret these clusters as the long-lived spots and measure the temporal evolution of the spot area in each cluster.
Most of the spots show very complex areal evolutions probably due to the consecutive flux emergence events in the same place, while some of them show clear emergence and decay phases as plotted in Figure \ref{fig:eachspot}.

We then compared the evolution of star spot areas detected by (a) local minima tracing method, (b) light curve modeling method, and (c) transit method in Figure \ref{fig:compmethod}.
We focus on the spot group A in Figure \ref{fig:lc:transit} to see how the temporal evolution of star spot areas computed by the different methods compare.
In panel (a) the spot group A indicated in Figure \ref{fig:lc:transit} is plotted.
In panel (b) the corresponding spot evolution estimated by light-curve modeling is plotted on the basis of the location in Figure \ref{fig:compmethod}.
 As we expected, both of them show very similar temporal evolutions. This is very natural because both are obtained from the same data, but important for the validation of the local minima tracing method. 
In panel (c) we plot the temporal evolution of the selected spot area estimated by the transit method which is located on the longitude between 0$^\circ$ and 100$^\circ$ in the bottom panel of Figure \ref{fig:lc:transit}.
 As one can see, the spot group A actually consists of at least four spots (two spots at the same time), and the temporal evolution of in-transit spots does not match with that of spot group A. 
The red circle shows the most dominant spot showing a clear emergence/decay phase, but the peak time is different from those seen in panel (a) and (b).
This means that the temporal evolutions of the individual spots are different from those of spot groups (active longitudes), especially in this period.

We estimated the temporal evolution of individual star spot areas occulted by the planet and compared with those derived by other methods.
We estimated the evolution of star spot area shown in Figure \ref{fig:eachspot} with linear fitting (see red lines in Figure \ref{fig:eachspot}), and calculated the maximum area (flux), emergence rate, decay rates, and lifetime in the same manner as \cite{2019ApJ...871..187N}.
We assume the mean magnetic field is about 2000 G when we compare them with the solar values.
The estimated emergence rate is $1.1\times10^{21}$ Mx$\cdot$h$^{-1}$ on average, and the decay rate is $-7.8\times10^{20}$ Mx$\cdot$h$^{-1}$, for the spot area of $1.3\times10^{24}$ Mx.
On the other hand, the emergence/decay rates of spot group derived from local minima is $3.8\times10^{20}$ Mx$\cdot$h$^{-1}$ and $-5.6\times10^{10}$ Mx$\cdot$h$^{-1}$ for the spot area of $2.0\times10^{24}$ Mx, respectively.
Likewise, the emergence/decay rates of spot group derived from light curve modeling is $1.1\times10^{20}$ Mx$\cdot$h$^{-1}$ and $-3.9\times10^{20}$ Mx$\cdot$h$^{-1}$ for the spot area of $1.4\times10^{24}$ Mx, respectively.
The spot area is quite similar, but the flux emergence/decay rates derived from rotational modulations are smaller by one order of magnitude than those derived from the transit method.
 Here, the spot group A has an equatorial rotational period and expected to be on round the equator, so the latitudinal effects are negligible. 
Each data are plotted in Figure \ref{fig:emergedecay} in comparison with those of sunspots and star spot in our previous work.
As you can see, however, the emergence/decay rates of star spot occulted by transit look consistent with sunspots and star spots in our previous work, although they scatter by an order of magnitude.
The emergence rates can be explained by the solar scaling relation \citep[d$\Phi$/d$t\propto\Phi^{0.3-0.5}$,][]{2011PASJ...63.1047O,2017ApJ...842....3N}, and the decay rates can be also explained by the solar relation \citep[d$\Phi$/d$t\propto\Phi^{\sim0.5}$,][]{1997SoPh..176..249P}.

We also compared the lifetime-area relation in Figure \ref{fig:lifetime}.
As a result, the lifetimes of star spots occulted by transit are consistent with previous star spot studies \citep{2014ApJ...795...79B,2015PhDT.......177D,2017MNRAS.472.1618G,2019ApJ...871..187N}, and the lifetimes of star spot are much shorter than extrapolated from the solar empirical relations \citep[$T\propto A$, so-called Gnevyshev-Waldmeier law;][]{Gnevyshev,1955epds.book.....W}.
 On the other hand, the area-lifetime relation is roughly derived to be $T$ [d] $\sim$ C ($A$ [MSH])$^{0.5}$, where C is $\sim 0.8$, from the above dependence of emergence/decay rates on the total flux.
This empirical relation would be more consistent with the stellar observations than the Gnevyshev-Waldmeier law as discussed in the following section.

\section{Discussion} \label{sec:dis}

\subsection{Spot Area}

In this section, we summarize and discuss the locations and spot areas on Kepler 17.
In Sect. \ref{res:1}, we showed that one local minimum actually consists of several dominant spots in the case of Kepler 17, which has been already indicated by \cite{2019arXiv190404489L}.
This can be easily understood in analogy with sunspot distributions where we can see several active regions at the same time during the maximum activity.
 The locations of local minima sometimes nicely match those of (nearly equatorial) transited dominant spots; in those cases we can pin much of the light curve modulations to those spots (see, Figure \ref{fig:validation_kepler17}). 

We also showed the amplitudes of local minima values also have a positive correlation with the visible maximum spot area in transit and the total projected spot area in transit in this observational period.
 This positive correlation may imply that (1) the unseen spots are randomly distributed and so have a relatively small effect on the brightness variation and the dominant spots (group) in transit mainly create the rotational modulation \citep[e.g.][]{1994ApJ...420..373E} or (2) the unseen spots follow the same general hemispheric pattern as the transited spots.
In the case of (2), the unseen spots are perhaps part of the same general active areas, but not always, because sometimes the areas trend away from each other.

We also showed that the amplitudes of local minima match the visible maximum spot area in transit, while they are smaller by a factor of 2 than the total projected spot area in transit.
This factor difference would be caused because spots are widely distributed in longitude (see Figure \ref{fig:dbscan}), which decreases the brightness variation amplitude as well demonstrated by \cite{1994ApJ...420..373E}.
 
The consistency between the amplitudes of local minima and the visible maximum spot area in transit indicates that the spot area derived from local minima (or amplitude of the brightness variations) can be still a good indicator of that of the largest spot on the disk, but do not always correspond to the total filling factor.
The similar properties can be seen in the case of the Sun as \cite{2018ApJ...865..142B} showed.


\subsection{Temporal Evolution of Individual Star Spot Areas}

In this section, we discuss the temporal evolution of star spot area in comparison with those of sunspots and our previous study \citep[][]{2019ApJ...871..187N}.
In Figure \ref{fig:emergedecay}, it is clear that the emergence/decay rates of spatially resolved star spot (resolution is still $\sim$ 10-20 deg) are consistent with those of sunspots  within one order of magnitude of error bars of solar data. 
Also, this result is roughly consistent with our previous study  within one order of magnitude.  \citep[red and blue circles,][]{2019ApJ...871..187N}, which measured the variations rates of the favorable individual spots (i.e., isolated local-minima series showing clear emergence and decay).
This  possibly supports  that the spot emergence/decay can be explained by the same mechanism, and imply a possibility to apply solar physics to star spot emergence/decay.
 However, the solar ($P_{\rm rot} \sim$ 25 d) and Kepler-17 ($P_{\rm rot} \sim$ 12 d) data is more consistent with the rapid rotators ($P_{\rm rot} <$ 7 d; blue circles) and larger than the slowly rotators  ($P_{\rm rot} >$ 7 d; red circles).
We speculate that the discrepancy between spatially resolved and non-resolved spots on the slowly rotating stars can be a result of the superposition effect of the several spot evolutions, as showed in Section \ref{res:3} and discussed in the next paragraph.
As for the rotational period dependence, \cite{2019ApJ...871..187N} have discussed some possible mechanisms of the dependence of the rotational period on the spot evolution (e.g. decay due to the differential rotations of the stars), although it is still debated.


How about the comparison between the in-transit  (spatially-resolved)  and rotational-modulation  (spatially-unresolved)  spots? 
In the Kepler-17 system, it is reported that  the two active longitudes are prominent,  whose lifetimes are over the $Kepler$ observational period ($\sim$ 1400 days), while the brightness variation amplitude is varying every moment \citep{2019arXiv190404489L}.
This is  partly  due to intermittent flux emergence and decay on the stellar surface.
In fact, in Sect. \ref{res:3}, we also showed that the temporal evolution of individual star spot area in transit looks different from those derived from the local minima tracing method and light curve modeling method.
These results would imply that the temporal evolution of star spot derived from out-of-transit light curves (i.e., rotational modulations) can be actually a superposition of the several dominant spots existing at the same active longitude.
As a result of the superposition effect, as in Figure \ref{fig:emergedecay}, the emergence/decay rates of the star spots in transits ($\sim\pm1\times10^{21}$ Mx$\cdot$h$^{-1}$) are larger by more than a factor of two (up to one order of magnitude) than those derived from out-of-transit light curves ($\sim\pm1$-$6\times10^{20}$ Mx$\cdot$h$^{-1}$), although the maximum spot area is quite similar ($\sim10^{24}$ Mx).
This is a feature discovered only on Kepler 17, but can be applicable to the other stars.
 So far, most of the star spot evolutions are estimated based on the $Kepler$/$COROT$ rotational modulations \citep{2014ApJ...795...79B,2017MNRAS.472.1618G,2019ApJ...871..187N}, our results propose that there is a certainty that the superposition effect changes the lifetime and variation rates by some factor, and the lifetimes may not be those of spot groups but those of active longitudes.
Even though \cite{2019ApJ...871..187N} carefully chose the favorable targets, we therefore have to be careful on the qualitative discussions.

 In fact, most of the in-transit spots do not show clear emergence/decay phase.
This may mean that there are continuous flux emergences and as a result they do not show clear emergence/decay phase, but can mean that the we may pick up only spots having rapid emergence/decay and the variation rates can actually have much larger diversity (more than one order of magnitude).

We also comment on the difference between the light curve modeling method and the local minima tracing method.
In Sect. \ref{res:3}, the temporal evolution of star-spot area derived by light curve modeling are consistent with those derived by the local minima tracing method, and they have similar values of the variation rates ($\sim\pm1$-$6\times10^{20}$ Mx$\cdot$h$^{-1}$) and the maximum spot sizes ($\sim10^{24}$ Mx).
The reasons for the small difference in variation rates (a factor of $\sim$ 5) are as follows.
First, the light curve modeling method improves the results of the local minima tracing method by considering the contamination of different spots.
Second, the $Kepler$ light curves have inevitable long-term trends, so the relative values between different observational quarters are not reliable.
In our light curve, the mean value is set to be zero for each quarter.
Because of this, the light curve modeling method can generate a pseudo-long-lived spots, which result in the above difference  \citep[see,][for more detais]{2018ApJ...865..142B}.


Finally, we comment on the star spot lifetime.
Figure \ref{fig:lifetime} shows that the lifetimes of detected star spots are smaller than those expected from solar empirical relation (T = A/D; D=10), and the result is consistent with the other studies.
This is not surprising because the positive correlations of the variation rates and spot area (d$A$/d$t\propto A^{\alpha}$, $\alpha$=0.3-0.5) result in more small power-law relations $T\propto A^{1-\alpha}$ (the detail about the star spot lifetimes are discussed by \cite{2019ApJ...871..187N}).
Please note that Figure \ref{fig:dbscan} that all of the individual star spots seem to appear for only several hundred days \citep[see also][]{2015PhDT.......177D,2019arXiv190404489L}. 
This fact can be a strong restriction on the spot-evolution physics because our previous study \cite{2019ApJ...871..187N} could not exclude the existence of individual spots with lifetimes of more than 1,000 days as predicted from the solar Gnevyshev-Waldmeier relation.

\subsection{Implication for the Stellar Superflares}

In the Kepler 17, we cannot find any stellar flares in the light curve, although the stellar brightness variations indicate the existence of large star spots that have a potential to produce a superflare ($>10^{34}$ erg).
Also, \cite{2017PASJ...69...41M} reported solar-type stars that do not show any superflares but have large star spots.
The reason why such stars with large star spots do not show any superflares is an open question.
Statistically, stars with such large star spots $>10^{-1.5}$ cause superflares 0.1-1 times per one year \citep{2017PASJ...69...41M}, so such stars without superflares during $Kepler$'s observational period can be classified to have relatively less flare-productive spots.
One possibility is that the large spots without any flares, like spots on Kepler 17, can have a simple polarity shape (e.g., $\alpha$-type or $\beta$-type spots), which is known to rarely produce extreme flares in the case of the sunspots \citep[e.g.,][]{2019arXiv190412027T}.
On sunspots, complex spots show relatively fast decay, so that lifetimes of star spots can be an indicator of spot complexity and flare productivity.
The comparison of star-spot lifetimes between flare-productive and less flare-productive stars would be important for why and how stellar superflares occur.

\cite{2018ApJ...868....3R} reported that the stellar superflares do not appear to be correlated with the rotational phase on solar-type stars.
They tried to explain this result by the existence of large polar spots, visible large flares over the limb, or the flares between the active longitudes.
Our transit model reveals that there are, more or less, visible spots on the disk regardless of its rotational phase, and the local maxima are not always the unspotted brightness level.
These observations imply that, even if we assume that the superflares are accompanied with the solar-like low-latitude spots, superflares can apparently occur regardless of its rotational phase because large spots are always visible to some extent.

Finally, large star spots having a potential to produce superflares are found to survive more than 100 days (up to 1 year). 
This means that the surrounding exoplanets can be exposed to the threat of stellar superflares such a long period once large star spots appear, which can be critical for the exoplanet habitability \citep[e.g.,][]{2010AsBio..10..751S,2016NatGe...9..452A,yamashikiflare}.

\section{Summary and Future Prospects} \label{sec:sum}

In this study, we investigated the temporal evolution of individual spot areas by using the local minima tracing method, light curve modeling method, and transit method.
By using the transit method, we can estimate the properties of the (partially) spatially resolved star spots.
Kepler 17 is one of the best targets to analyze the temporal evolution of star spots by the transit method.
The main results in this study are as follows. 
(i) On Kepler-17, although two series of local minima are prominent based on the rotational light curve, there are clearly many spots present on the star based on the exoplanet transits. 
The location, area, and temporal evolution of one local-minima spot does not correspond to those of in-transit spots.
This means that we have to be careful when we derive the spot information based on the rotational modulations.
(ii) Nevertheless, the estimated area from the local minima tracing method is consistent with the maximum in-transit spots, indicating that the \textit{Kepler} light curve amplitude is a good indicator of the maximum visible spot size.
(iii) Although the temporal evolution derived from the rotational modulation differs from those of in-transit spots to a certain degree, the emergence/decay rates of in-transit spots are within an order of magnitude of error bars of those derived for sunspots.
This consistency possibly supports the possibility of applications of sunspot emergence/decay physics \citep[e.g.,][]{2011PASJ...63.1047O,2017ApJ...842....3N,1997SoPh..176..249P} to star spot evolutions.
It is also consistent with that based on rotational modulations \citep[e.g.][]{2019ApJ...871..187N} within one-order of magnitude, but slightly different for slowly rotating stars.
This would be because the evolution of local minimum is a superposition of that of a few large  spots, which produces a difference between spatially-resolved and spatially-unresolved star spot evolution.
(iv) This may not be surprising, but the star spot  distribution  derived by the light curve modeling method is well consistent with that of local minima tracing method in terms of spot location and area, implying that even the simple local minima tracing method can capture the  essential  feature of the rotational modulation. Although this kind of approach can fit the rotational light curves, the fitted results can be largely different from the real distribution of star spots. 

The above results are valid only for Kepler 17, and it's not obvious that these are applicable to the other stars. 
Kepler 17 is a solar-type star with the medium rotation period ($P_{\rm rot}$ $\sim$ 12 days),
and main physics to determine the spot emergence and decay can be different
from those of younger more rapidly rotating star ($P_{\rm rot}$ $\sim$ a few days) and older slowly-rotating stars like the Sun ($P_{\rm rot}$ $\sim$ 25 days), as suggested by \cite{2019ApJ...871..187N}.

Therefore, the validations on other stars would be necessary for further universal understandings of the star spot physics.
Up to present, the number of good targets is quite limited in the \textit{Kepler} field because this kind of research requires suitable inclinations of the stars and planets and high-time cadence data.
In near future, \textit{TESS} \citep[Transiting Exoplanet Survey Satellite,][]{2015JATIS...1a4003R} will provide us mid-term (27 days - 1 year) stellar photometric data, and is expected to find many transiting exoplanets with two-minute cadence.
This may be good candidates to confirm our results on other stars, not only solar-type stars but also cooler stars. 

Currently, we hardly observe the magnetic flux configurations of spots below the photosphere, even on the Sun, although the local helio-seismology has a potential to estimate them.
As a future work, comparisons with numerical simulations on spot evolution would be important.
Recently, calculations to investigate spot evolution have been widely carried out \citep[e.g.,][]{2008ApJ...687.1373C,2014ApJ...785...90R}, but the evolution in the limited numerical box are still influenced by the initial and bottom-boundary conditions.
Numerical simulations of spot evolution covering from the convection zone have been performed \citep[][]{2019SciA....5.2307H,2019ApJ...886L..21T}, which may reveal the spot emergence and decay mechanism.




\bigskip
\textit{Kepler} was selected as the tenth Discovery mission.
Funding for this mission is provided by the NASA Science Mission Directorate.
The data presented in this paper were obtained from the Multimission Archive at STScI.
This work was also supported by JSPS KAKENHI Grant Numbers
JP15H05814, 
JP16H03955, 
JP16J00320, 
JP16J06887, 
JP17H02865, 
JP17K05400, 
and JP18J20048.





\clearpage

\begin{figure}[htbp]
\begin{center}
\includegraphics[scale=0.5]{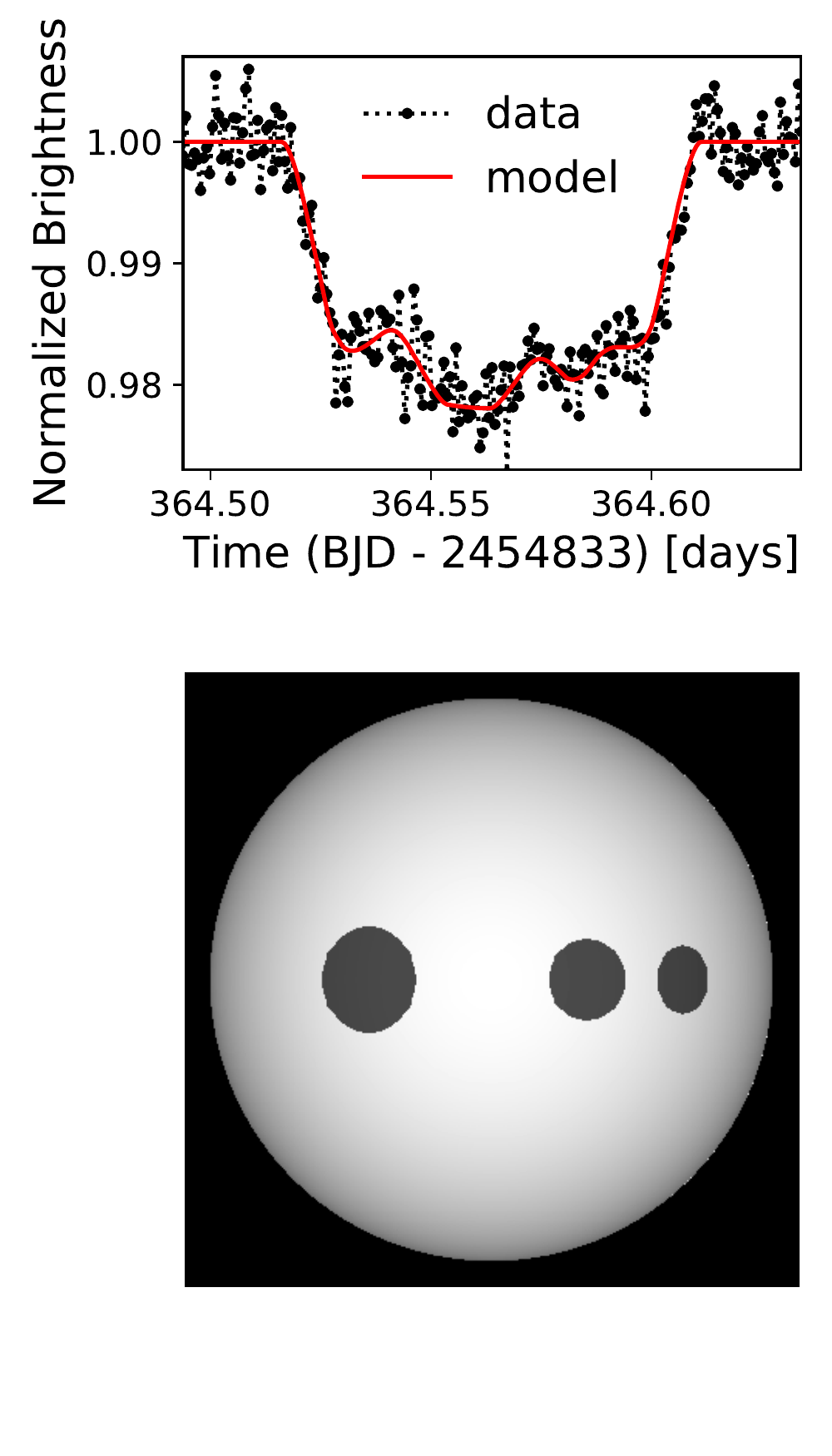}
\caption{Upper panel is one example of the high quality fits of our MCMC modeling (red line) to the high cadence in-transit data (black lines). Bottom panel is the reconstructed surface distribution of star spot.}
\label{fig:}
\end{center}
\end{figure}



\begin{figure*}[htbp]
\begin{center}
\includegraphics[scale=0.5]{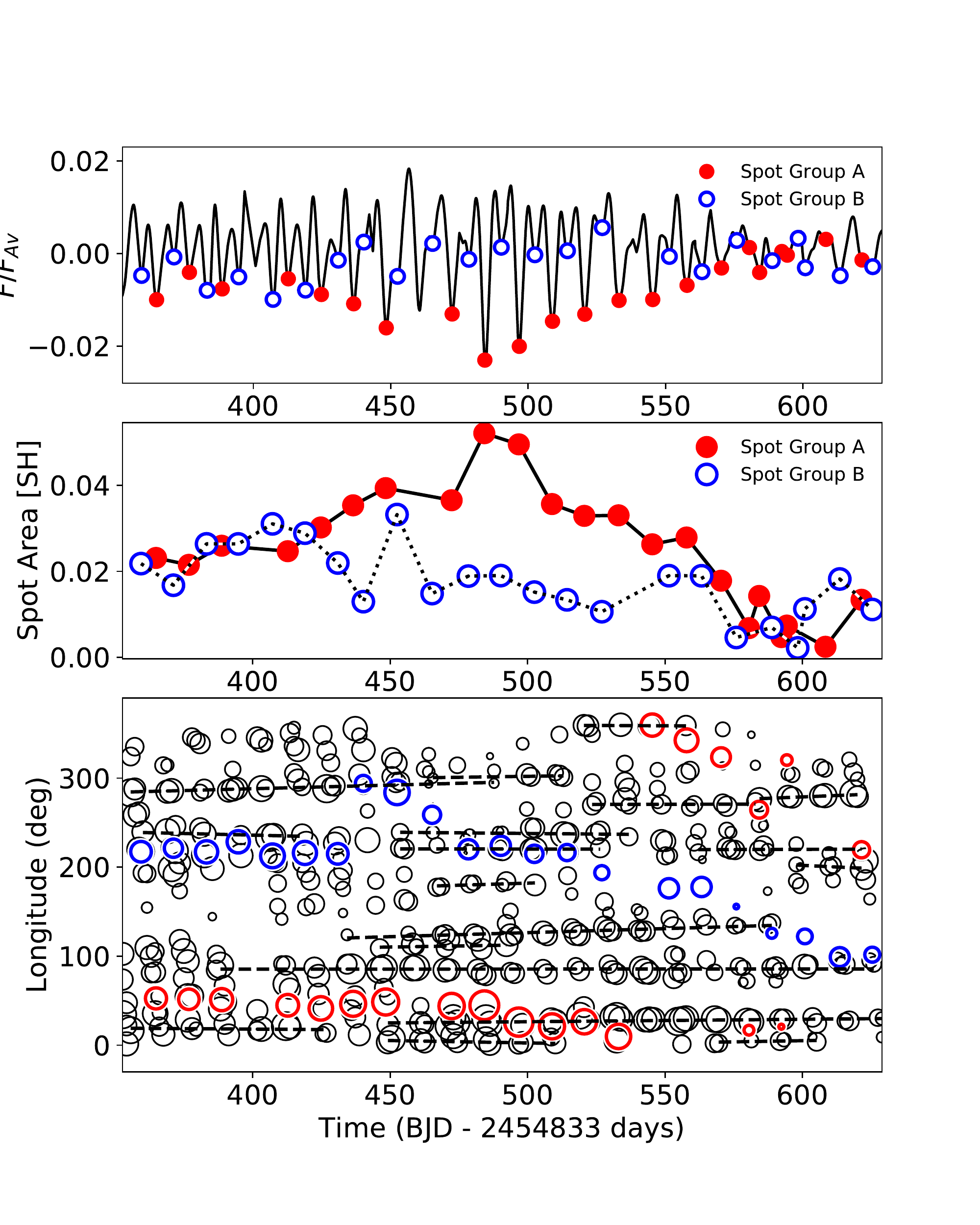}
\end{center}
\caption{Upper: the black line is the smoothed light curve of Kepler 17 during quarter 4--6. Circles indicate the local minima of the light curve, and the different symbols mean the different spot groups identified based on the rotational phase. Middle: the temporal evolution of the star spot area for each spot group. Bottom: the comparison on the spot distribution between in-transit spots (black) and local minima spots (colors). The longitude are calculated as the Carrington longitude with the rotational period of 11.92 days \citep{2017ApJ...835..294V}.  The dashed lines are the series of long-lived recurrent spot candidate identified by the {\sf DBSCAN} package in $python$.  }
\label{fig:lc:transit}
\end{figure*}

\begin{figure*}[htbp]
\begin{center}
\includegraphics[scale=0.5]{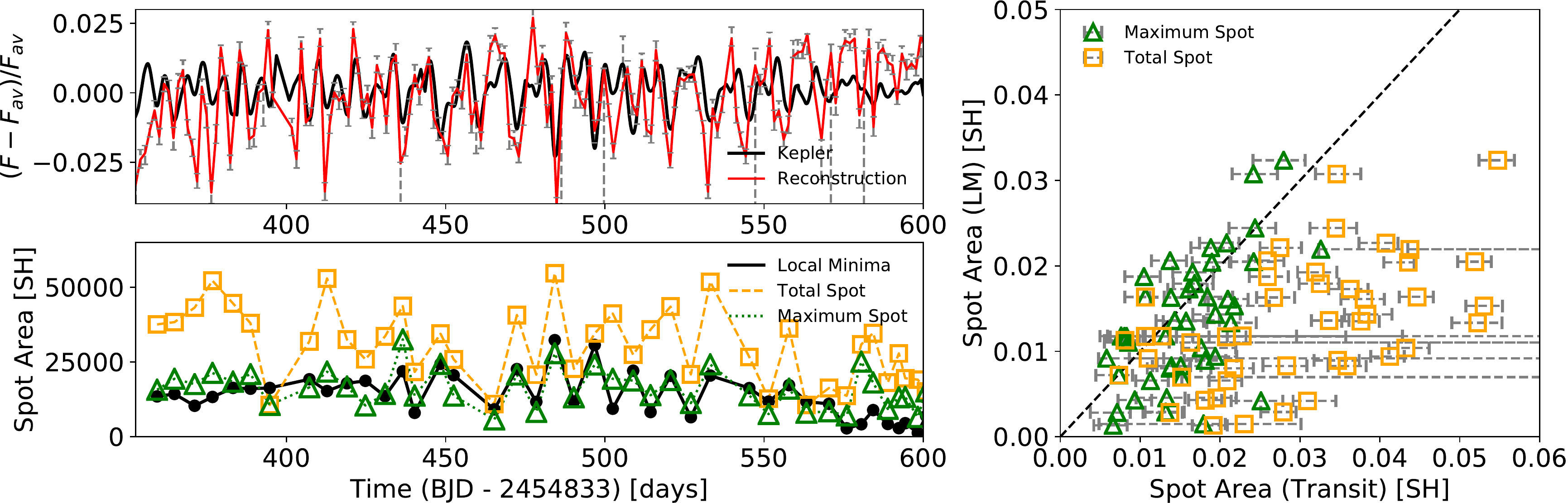}
\end{center}
\caption{Left upper: the comparison between the smoothed light curve of Kepler 17 (black) and reconstructed light curve by spots occulted by transits (red). 
 In this figure, the vertical axis values are those relative to their averaged values. The \textit{Kepler} light curve is subtracted by the average flux values for each quarter, while the reconstructed light curves are subtracted by the average value ($F_{\rm av}$) for all period, so both of them do not necessary match with each other. 
The error bars are estimated based on the posterior distributions of the spot area.
A zero-level value for the reconstructed light curve is $\sim 0.028$ which is slightly higher than the local maxima.
Left lower: temporal evolution of the spot area derived  by local minima tracing method (black), total spot area derived by transit method (orange), and maximum spot area derived by transit method (green). Right: comparisons between spot area derived by local minima tracing method and total spot area derived by transit method (orange), and comparisons between spot area derived by local minima tracing method and maximum spot area derived by transit method (green).}
\label{fig:validation_kepler17}
\end{figure*}

\begin{figure*}[htbp]
\begin{center}
\includegraphics[scale=0.5]{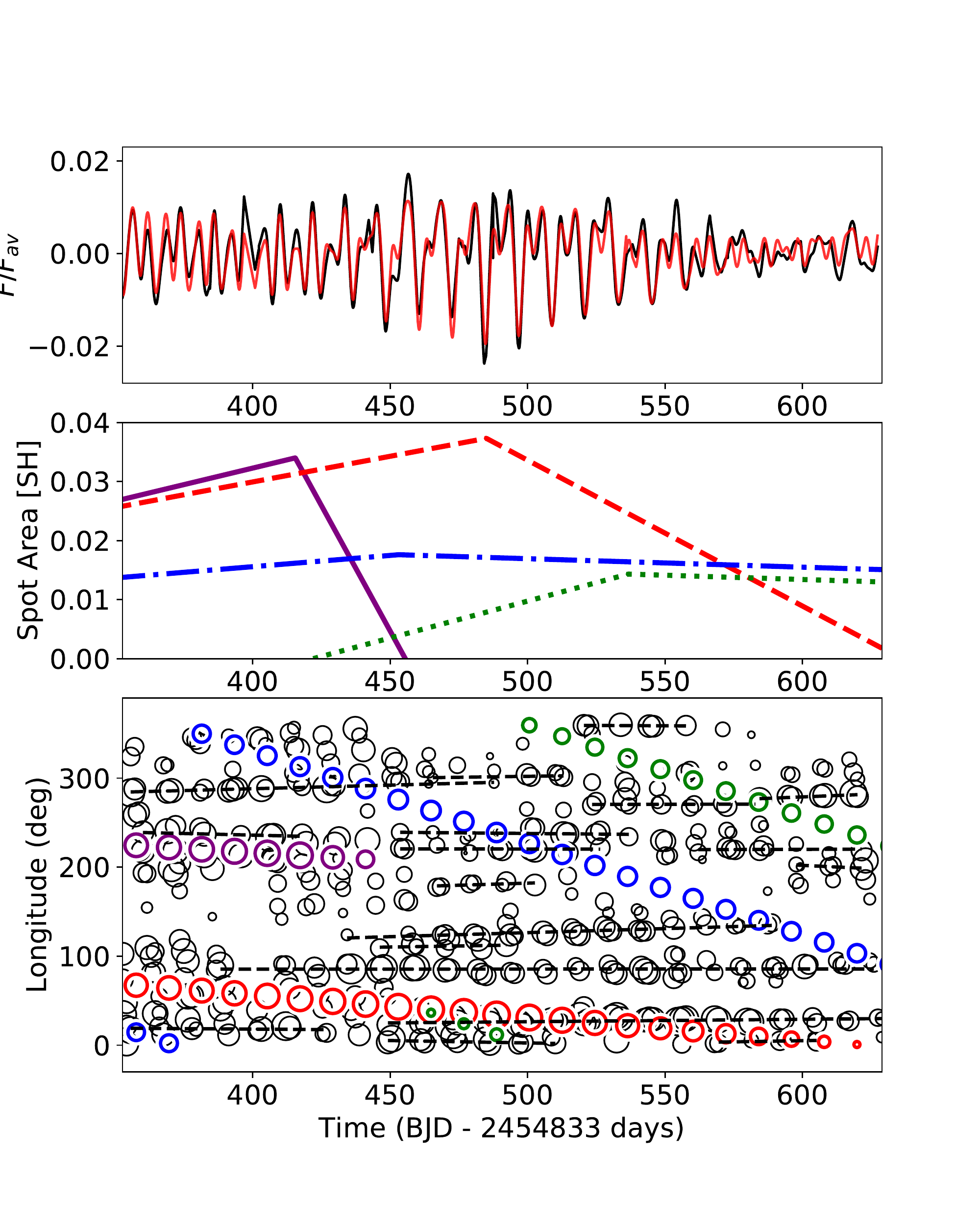}
\end{center}
\caption{Upper: the black line is the smoothed light curve of Kepler 17 during quarter 4--6, and the red line is the modeled light curve with four-spots light curve modeling method. Middle: the temporal evolution of the star spot area estimated in our model. Bottom: the comparison on the spot distribution between the spots occulted by the transit and the those estimated by light curve model.  The dashed lines are the series of long-lived recurrent spot candidate identified by the {\sf DBSCAN} package in $python$.  }
\label{fig:ReMCSTSP}
\end{figure*}

\clearpage

\begin{table*}\label{table:2}
\begin{center}
\scriptsize
\begin{tabular}{lcccccc}
\hline
 & Maximum Area & Emergence Rate & Decay Rate & Rotation Period  \\
 & [MSH] & [MSH$\cdot$h$^{-1}$] & [-MSH$\cdot$h$^{-1}$] & [d] \\
\hline
Spot 1 & 20600$^{+1.1}_{-200}$ &2.83$^{+0.47}_{-0.21}$ &21.4$^{+0.23}_{-0.52}$ &11.99$^{+0.0047}_{-0.00045}$  & \\ 
Spot 2 & 22600$^{+130}_{-36}$ &2.2$^{+0.48}_{-0.39}$ &6.23$^{+0.084}_{-0.017}$ &12.02$^{+0.0048}_{-0.0034}$  & \\ 
Spot 3 & 8670$^{+91}_{-65}$ &3.16$^{+0.055}_{-0.01}$ &0.363$^{+0.028}_{-0.0024}$ &12.34$^{+0.017}_{-0.016}$  & \\ 
Spot 4 & 10700$^{+120}_{-240}$ &0.97$^{+0.15}_{-0.26}$ &0.362$^{+0.00063}_{-0.0018}$ &12.34$^{+0.017}_{-0.016}$  & \\ 
\hline
\end{tabular}
\caption{The spot parameters estimated by our MCMC-based light curve modeling for the rotational modulation. The values are taken as parameter showing the highest likelihood, and the error bars are estimated from 68 \% of the posterior distribution.}
\end{center}
\label{tab:mcmc}
\end{table*}

\clearpage

\begin{figure}[htbp]
\begin{center}
\includegraphics[scale=0.5]{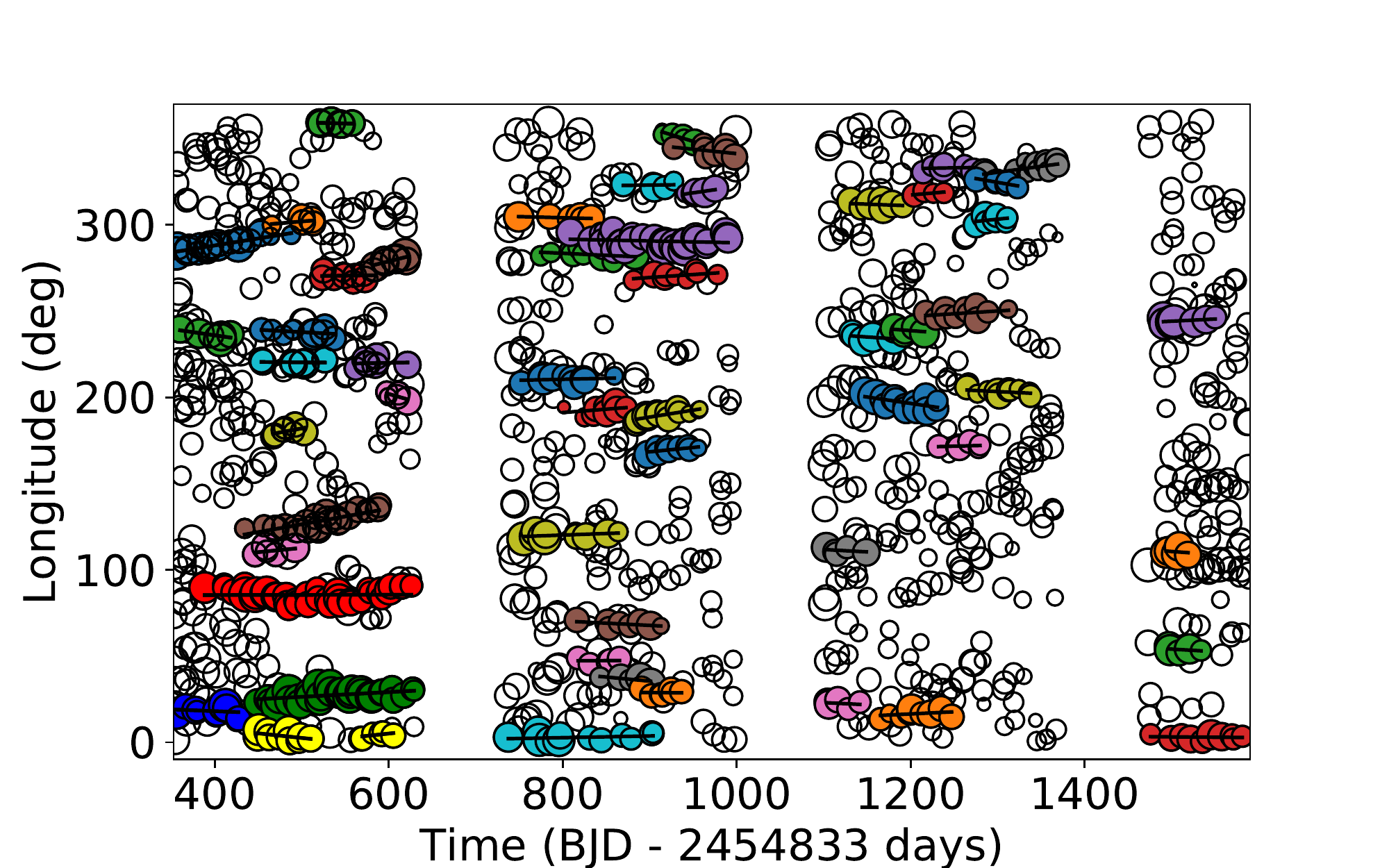}
\end{center}
\caption{Symbols shows all spots detected by the {\sf STSP} code \citep{2015PhDT.......177D,2017ApJ...846...99M} on the time-longitude diagram. The color symbols are the series of long-lived recurrent spot candidate identified by the {\sf DBSCAN} package in $python$.}
\label{fig:dbscan}
\end{figure}

\begin{figure}[htbp]
\begin{center}
\includegraphics[scale=0.5]{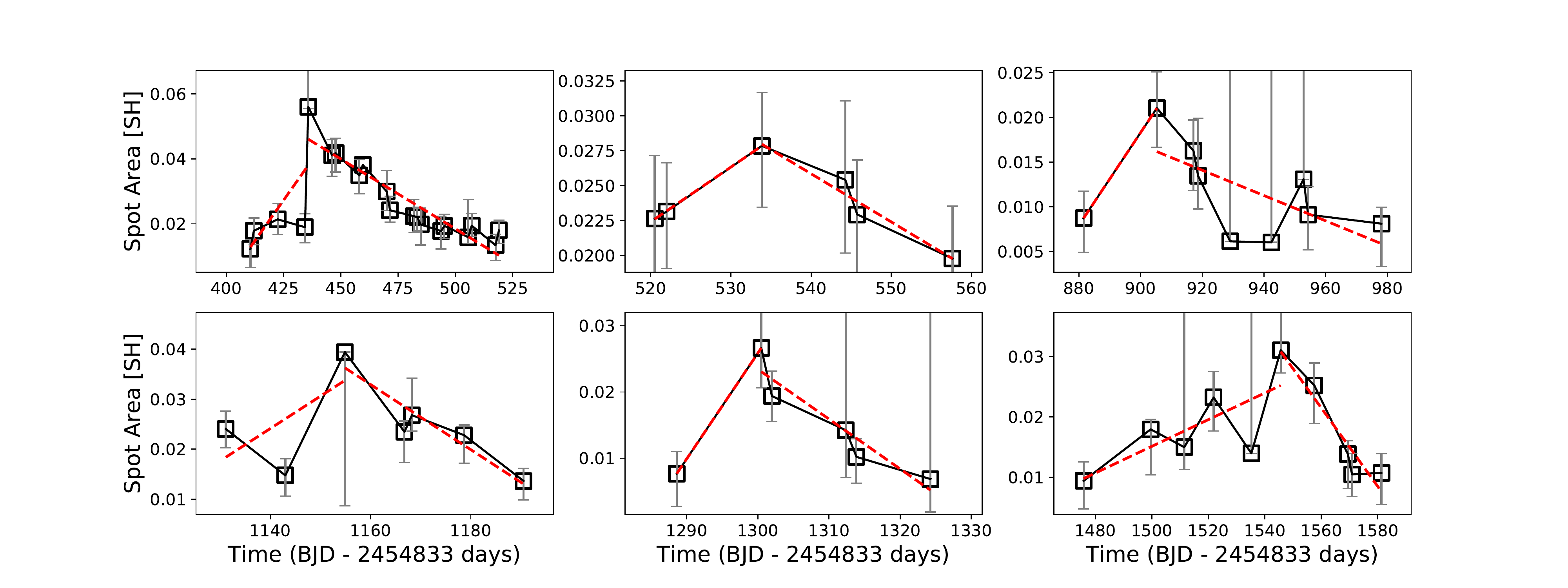}
\end{center}
\caption{Several examples of the temporal evolution of star spot estimated from the exoplanet transit model which show clear emergence and decay phase.  The solid error bars are estimated from 68 \% of the posterior distribution.    }
\label{fig:eachspot}
\end{figure}

\begin{figure}[htbp]
\begin{center}
\includegraphics[scale=0.6]{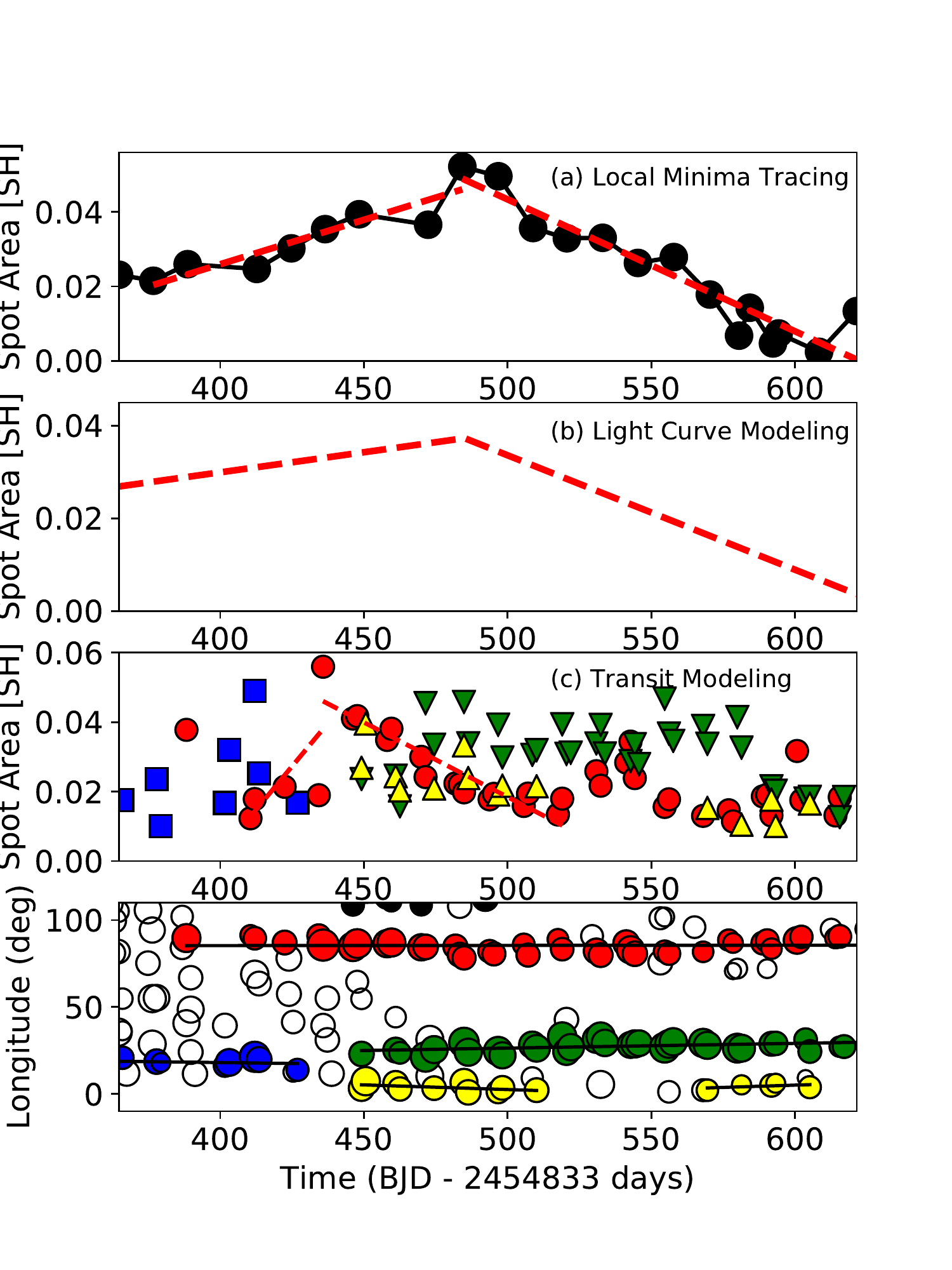}
\end{center}
\caption{The temporal evolutions of star spot area derived by (a) local minima tracing method, (b) light curve modeling method, and (c) {\sf STSP} transit method. In panel (a), the emergence and decay phases are fitted and indicated with the red dashed line. In panel (b), the corresponding star spots are indicated by red lines. In panel (c), 
 the derived areas of the spots in the first observing period between longitude 0$^\circ$ and 100$^\circ$ are shown; the primary growth and decay of the spot near 100$^\circ$ is marked with the red dashed line.  
In the bottom panel, we show the extended figure of Figure \ref{fig:dbscan} whose longitude is between 0$^\circ$ - 100$^\circ$. }
\label{fig:compmethod}
\end{figure}


\begin{figure}[htbp]
\begin{center}
\includegraphics[scale=0.45]{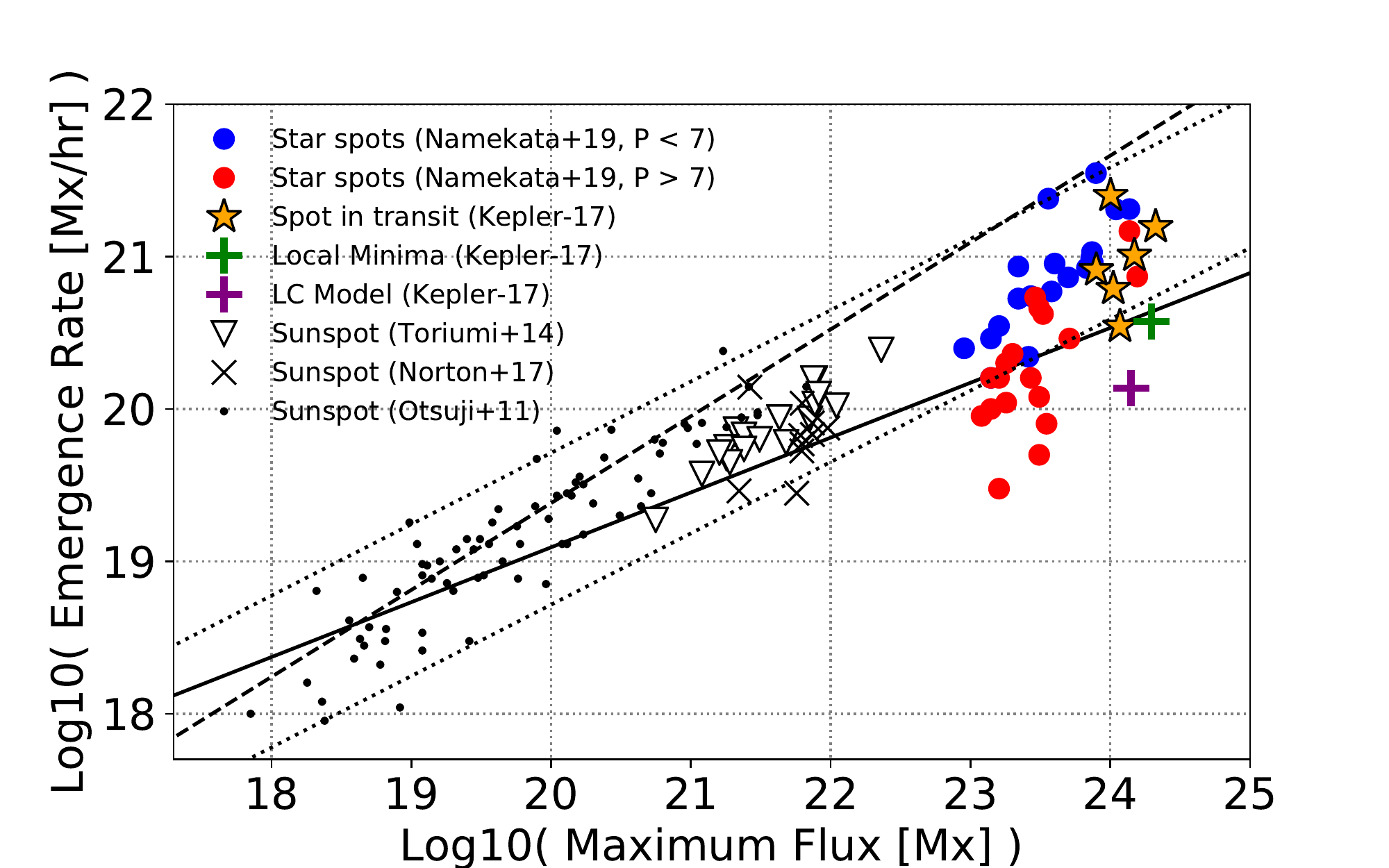}
\includegraphics[scale=0.45]{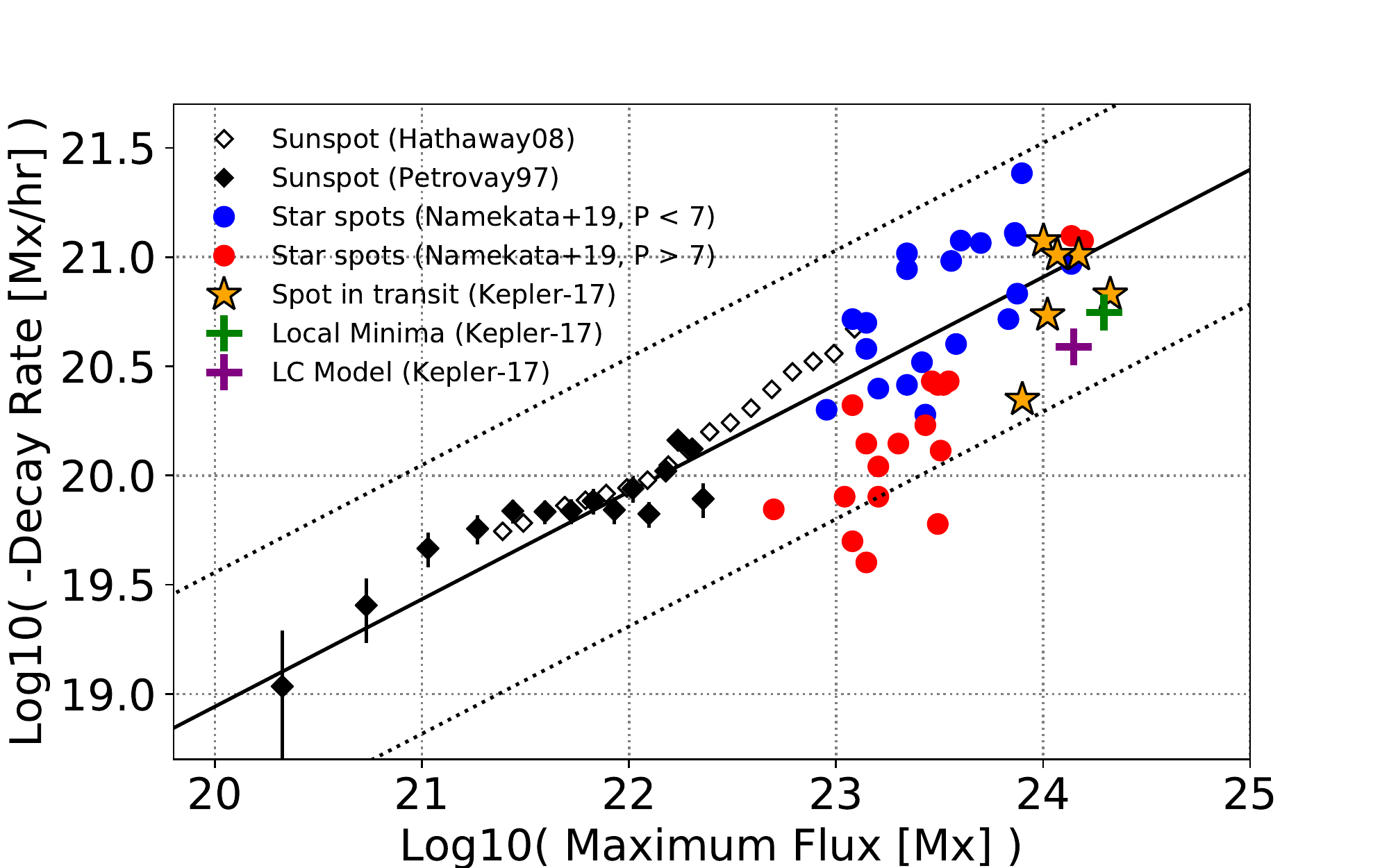}
\end{center}
\caption{Left: comparison between maximum magnetic flux and emergence rate of sunspots and star spots on solar-type stars. Black points, downward triangles, and crosses are sunspots observational data taken from \cite{2011PASJ...63.1047O}, \cite{2014ApJ...794...19T}, and \cite{2017ApJ...842....3N}, respectively. Blue and red circles correspond to the spots analyzed in \cite{2019ApJ...871..187N} with Prot $<$ 7 day and Prot $>$ 7 day, respectively. The yellow star symbols indicates the star spot occulted by transit derived in this study. A solid and dashed line is a scaling relation derived by \cite{2017ApJ...842....3N} and \cite{2011PASJ...63.1047O}, respectively.   The dotted lines are the 95 \% confidence level of solar data taken from \cite{2019ApJ...871..187N}.   Right: Comparison between maximum magnetic flux and decay rate of sunspots and star spots on solar-type stars. Black open and filled diamonds are sunspot's observations by \cite{2008SoPh..250..269H} and \cite{1997SoPh..176..249P}, respectively. Blue and red circles correspond to the spots analyzed in \cite{2019ApJ...871..187N} with Prot $<$ 7 day and Prot $>$ 7 day, respectively. A solid line is the line of the power law index of 0.5, where the absolute values are derived based on mean values of the sunspot observations \citep{2008SoPh..250..269H}. The yellow star symbols indicates the in-transit star spots derived in this study. The green and purple crosses indicates the star spot on Kepler 17 derived by using local minima tracing method and light curve modeling method  for the red lines in Figure \ref{fig:compmethod} (a) and (b),  respectively.   The dotted lines are the 95 \% confidence level of solar data taken from \cite{2019ApJ...871..187N}.  }
\label{fig:emergedecay}
\end{figure}

\begin{figure}[htbp]
\begin{center}
\includegraphics[scale=0.5]{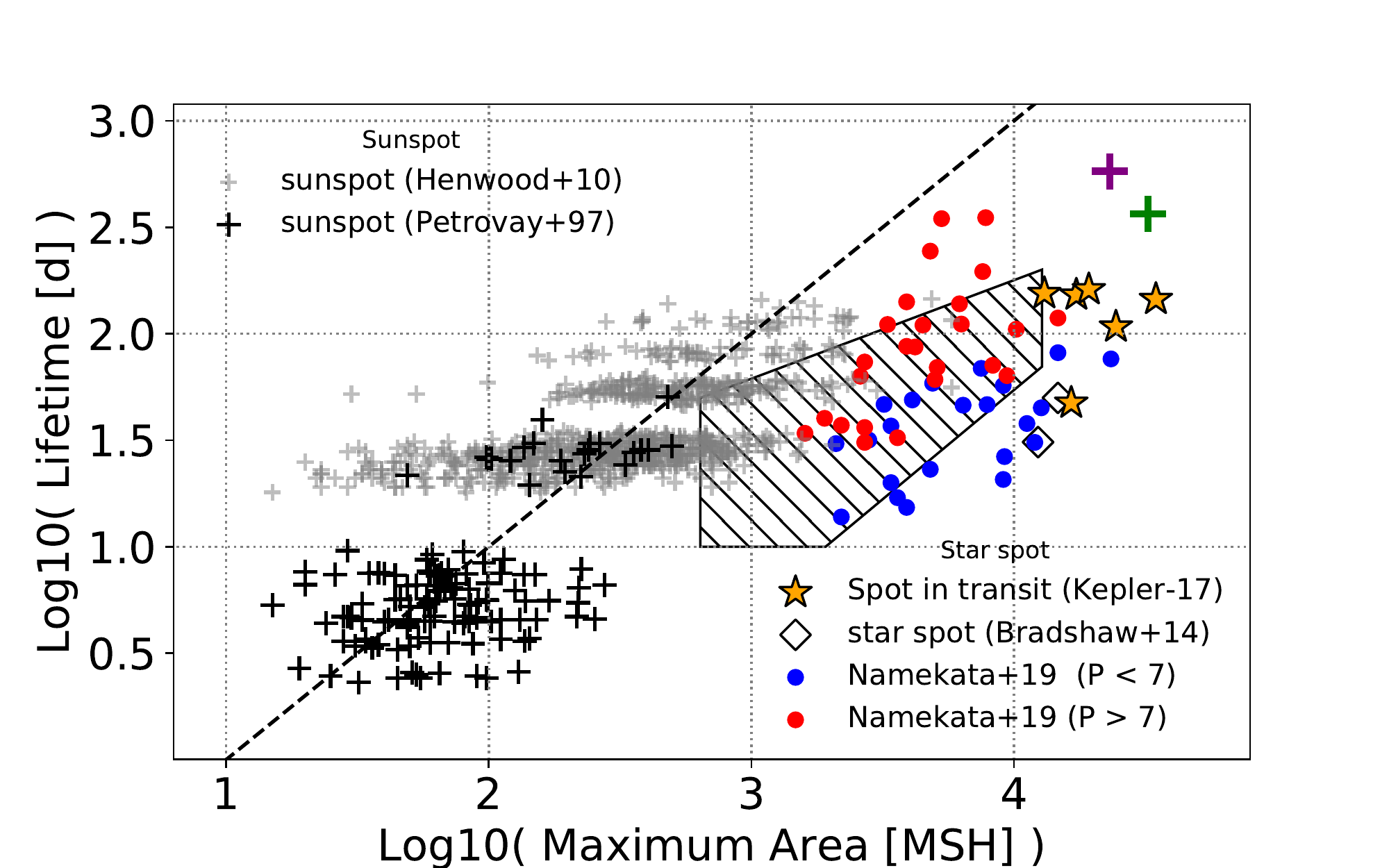}
\end{center}
\caption{Comparison between maximum spot area and lifetime of sunspots and star spots on solar-type stars. Black and gray crosses are sunspots data taken from \cite{1997SoPh..176..249P} and \cite{2010SoPh..262..299H}, respectively. The dashed line indicates the solar GW relation (A = DT, D =10 MSH/day). Blue and red circles correspond to the spots analyzed in \cite{2019ApJ...871..187N} with Prot $<$ 7 day and Prot $>$ 7 day, respectively. Open diamonds are star spots on G-type stars (Kepler 17 and CoRoT 2) taken from \cite{2014ApJ...795...79B}. A region filled with diagonal lines indicates the result of \cite{2017MNRAS.472.1618G}. The star symbols indicates our data in Kepler 17.  The green and purple crosses indicates the star spot on Kepler 17 derived by using local minima tracing method and light curve modeling method, respectively.}
\label{fig:lifetime}
\end{figure}

\clearpage

\appendix

\section{Extended Figure 4 for the other observational periods} \label{sec:app1}

\begin{figure}[htbp]
\begin{center}
\includegraphics[scale=0.5]{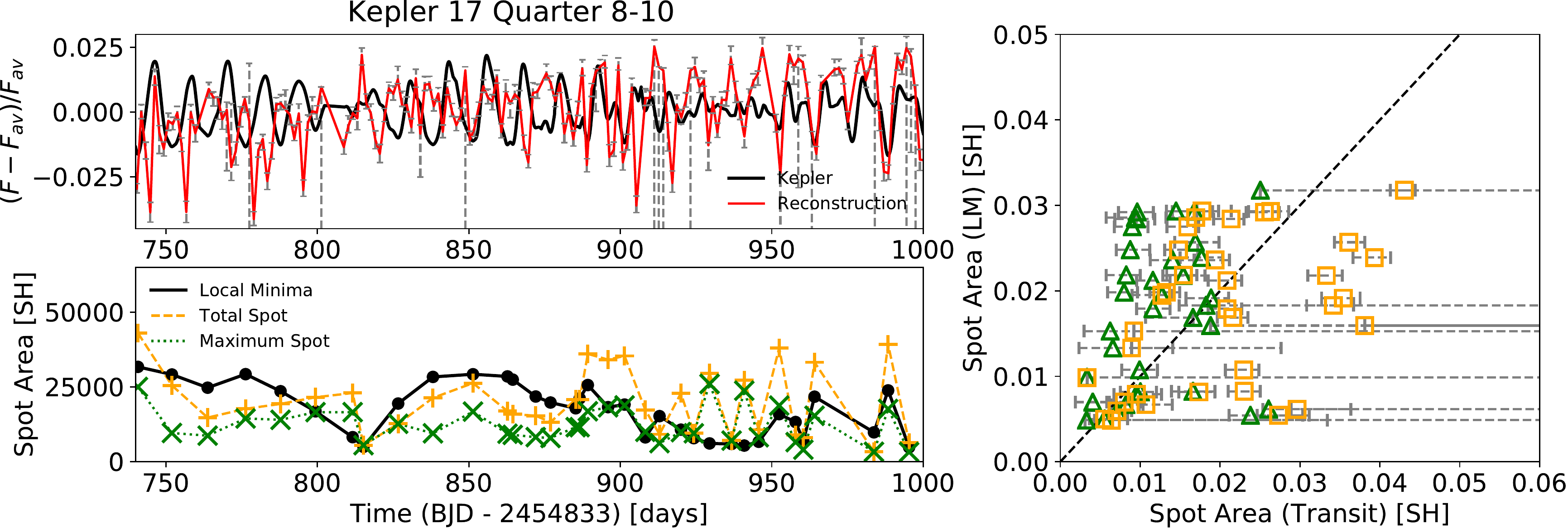}
\end{center}
\caption{This figure is basically the same as Figure 4, but for Quarter 8-10.}
\label{fig:app1}
\end{figure}

\begin{figure}[htbp]
\begin{center}
\includegraphics[scale=0.5]{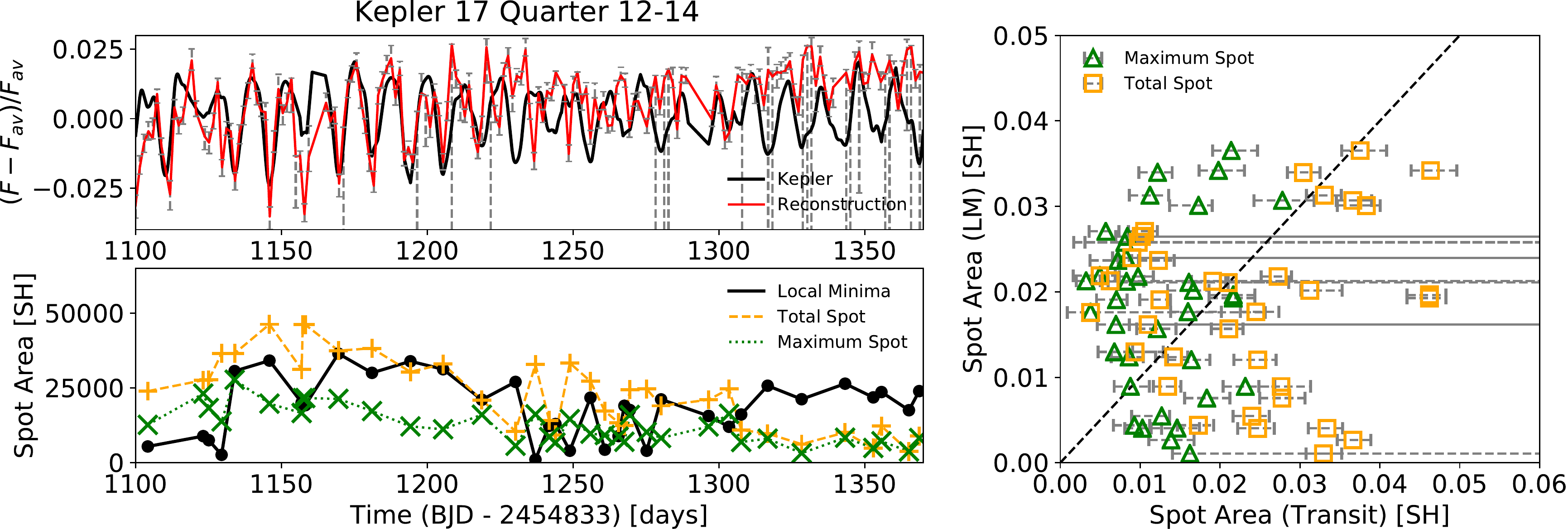}
\end{center}
\caption{This figure is basically the same as Figure 4, but for Quarter 12-14.}
\label{fig:app2}
\end{figure}

Figure \ref{fig:app1} and \ref{fig:app2} show the same analyses as Figure \ref{fig:validation_kepler17} for Kepler quarter 8 to 10 and 12 to 14, respectively.
The reconstructed light curve is sometimes consistent with but almost inconsistent with the Kepler-30-min light curve.
Also, we cannot see positive correlations between in-transit spot area and rotational-modulation spot area.
As \cite{2019arXiv190404489L} and \cite{2015PhDT.......177D} reported, the period of the rotational modulations are slower than that of the equator of Kepler 17 in later quarters, indicating that dominant spots exist out of the transit path.

\end{document}